# Institutional cooperations in Austrian research: An analysis of shared researchers

Christoph Schlager[1], Lutz Bornmann[2], Gerald Schweiger[1]

[1]Center for Artificial Intelligence and Machine Learning, Vienna University of Technology Vienna, Vienna, Austria

[2]Science Policy and Strategy Department, Administrative Headquarters of the Max Planck Society, Munich, Germany

Email:
Christoph Schlager – christoph.schlager@tuwien.ac.at
Lutz Bornmann – bornmann@gv.mpg.der
Gerald Schweiger – gerald.schweiger@tuwien.ac.at

**Abstract**

Multiple organisational affiliations are an increasingly common feature of research systems, yet their implications for organisational performance had received limited systematic attention. We developed a scalable, network-based analytical framework that represents simultaneous researcher affiliations as relational links between organisations and applied it to bibliometric data from Austria. Using harmonised publication and affiliation metadata, we constructed two complementary co-affiliation networks: a complete network capturing all simultaneous affiliations and a temporally filtered network retaining only organisational pairs that recurred over time. Analysing these networks in parallel allowed us to distinguish between transient and persistent co-affiliations and to assess how short-lived and enduring organisational ties shaped intra- and inter-organisational connectivity. Network regression analyses showed that geographical proximity remained an important determinant of co-affiliation formation, with spatial distance consistently reducing shared appointments. Clear sectoral differences emerged beyond geography. Universities formed a dense and persistent core of co-affiliations, whereas ties involving medical institutions, government, non-profit and private-sector organisations were often short-lived and attenuated under temporal filtering. Among cross-sector links, co-affiliations between universities and research institutes were notably resilient, indicating a more structurally embedded form of organisational integration. We assessed the effect of concurrent affiliations on organisational citation impact across organisational types using field- and year-normalised indicators. Sensitivity analyses showed that estimated impact levels and rankings depend on how concurrent affiliations were treated; however, the relative ordering of top organisations remained stable. Research institutes and universities consistently exhibited higher citation impact than organisations from other sectors, and persistent co-affiliations were associated with greater and more stable scientific visibility. These findings highlight the importance of explicitly modelling multiple simultaneous organisational affiliations: These affiliations exercise a substantial influence on scientific performance measures. The findings demonstrate that the scalable network-based framework (developed in this study) can yield detailed insights into the structure, persistence and performance implications of organisational ties in contemporary research systems. We recommend follow-up studies that address the topic of multiple simultaneous organisational affiliations in other countries using this framework.

# 1  Introduction

Research collaboration is recognized as a "salient feature of contemporary science" [1]. Recent literature has emphasized the importance of intra- as well as inter-institutional collaborations for academic research, which has been documented by the increase in team sizes and cross-institutional collaborations on co-authored papers [2]. In this academic environment that promotes collaboration and mobility, multiple affiliations are "increasingly recognised as facilitating knowledge exchange … and have been said to become more widespread" [2, p. 286].

Multi-affiliation refers to the practice of authors including multiple institutional affiliations in their publications, reflecting their connections to various organizations [3]. The traditional definition of institutional collaboration is narrow: "Institutional collaboration is commonly defined as authors being affiliated with at least two institutions in a coauthored article" [4, p. 550]. This definition implies that "more than one author and at least two institutions are the requirements for coauthored articles generated from interinstitutional collaborations" [4, p. 550]. The phenomenon of multiple affiliations challenges this strict definition, particularly when only a single author is involved, such as when "an individual researcher has two institutional affiliations – for example, a university department and a hospital – and these institutions are located in different countries" [5, p. 11]. Although "one could argue that there is no collaboration at all" [5, p. 12] in single-author cases, such publications are often viewed as valid institutional collaborations because they "will often reflect an agreement between departments or institutions to share a researcher" [5, p. 13]. The concept of the 'shared researcher phenomenon' suggests the solution is "to distinguish inter-institutional collaboration from inter-individual collaboration" [5, p. 16]. In this interpretation, the cooperation is manifested as a publication from the 'shared' researcher who lists the various institutional addresses. Therefore, "an indication of multiple affiliations by a single author in the address field nevertheless reflects an inter-organizational collaboration" [1].

The existence of multiple affiliations stems from a variety of motives, which can be categorized as individual or institutional [2,6]. At the individual level, multiple affiliations are often driven by personal research trajectories and professional development [7]. A key driver is the pursuit of resources and networking opportunities, as "affiliation to an institution is closely linked to resource access, research infrastructure and career opportunities" [2, p. 286]. Co-affiliations can offer direct and indirect access to research resources, including funding and equipment, which is critical in a science system reliant on extensive research infrastructure [6]. A prestigious affiliation can serve as a "mechanism for cumulative advantage" [8, p. 10729]. Individual motives for multiple affiliatios include the desire to stay connected with former institutions, especially for internationally mobile researchers [7,9]. These researchers may serve as 'bridging persons', coordinating research and exchanging knowledge between heterogeneous organizations [1,10]. Monetary or financial considerations can also motivate academics, particularly "more senior academics who seek to monetise on their reputation or expertise" [6, p. 385]. At the institutional level, competition and performance assessment strongly drive the practice of offering multiple affiliations [2]. Performance assessments have led universities globally "to become more proactive in attracting the most prolific researchers in a bid to enhance their position in national and international rankings" [2, p. 286].

Co-affiliations, which result in publications listing two or more institutions, are "likely high in research focus" [6, p. 384]. While approximately half of these affiliations are between academic institutions, non-academic organizations, including private firms, governments, and Non-Governmental Organizations (NGOs), are also frequently named [6]. Beyond research, multiple affiliations may serve purposes related to innovation or the 'third mission', including advisory tasks, managerial tasks, teaching purposes, and traditional honorary positions [6]. The prevalence of multiple affiliations in the science system has generally increased substantially over the past decade [7,11]. In terms of publication volume, on average, "about one quarter of articles" [7, p. 1044] qualify as having multiple affiliations. This figure rose from a low of 14.7% in 2001 to 32.5% in 2018 [7]. By 2019, "almost one in three articles was (co-)authored by authors with multiple affiliations" [7, p. 1039]. University affiliations are the most common institutional type involved [7]. The increase in multiple affiliations among researchers is related to the growing importance of bibliometric indicators for research funding distributions [2]. The increase can also be linked to the introduction of more competitive funding structures, such as 'excellence initiatives' (ExIns), which encourage universities to "buy in external talent instead of building up capacity locally or restructure their activities" [2, p. 1040].

Publications that contain multi-affiliated authorship have generally been found to have a "larger probability of receiving more citations" [10]. For example, a study using a large-scale analysis of scientific publications from four multi-disciplinary science journals (*Science*, *Nature*, *Proceedings of the National Academy of Sciences*, *PLOS Biology*) found a positive effect on citation counts [12]. The general pattern observed is that authors with multiple affiliations are "more often found on high impact papers" [2, p. 285] particularly concerning authors from Japan and Germany in biology and chemistry. Furthermore, researchers with multiple affiliations are "more often found in highly cited publications … implying their positive influence on scientific impact" [10]. Multiple affiliations can significantly benefit institutional performance: "Institutional academic performance benefits from multi-institutional authorship because a single author's publications contribute to the performance of more than one institution" [4, p. 550]. However, the influence of multiple affiliations on citation impact varies across disciplines. In some fields, multi-institutional authors show greater academic impact than other authors, but the opposite may occur in others [4]. For instance, "the academic impact of articles by multi-institutional authors was observed to be greater than that of other articles in high-energy physics (12.6 vs. 7.62 mean citations per article)" [4, p. 549], while "the opposite was observed in genetics (73.14 vs. 75.63 mean citations per article)" [4, p. 549]. When differentiating between national and international multi-affiliated authorship, studies have revealed differing patterns across disciplines [10]. It was found that in medicine-related and biology-related disciplines, "the national multi-affiliated authors are associated with more citations than the international multi-affiliated author" [10]. Conversely, space science, geosciences, and mathematics "show up the opposite phenomenon" [10], where "the international multi-affiliated authors relate to more citations" [10]. The study used a dataset of collaborative publications co-authored by two or more institutions published between 2013 and 2015, using a 3-year citation window for counting citations [10]. One limitation of this study may be that a short citation period (3-year citation window) was considered, suggesting that "the effect of multi-affiliated authorships on long-term citations needs further investigation" [10].

From a research evaluation perspective, multi-affiliation constitutes a particularly thorny methodological problem [9]. The whole counting method, which is still the predominant way of tallying papers and citations received by an institution or country, dictates that each author's affiliation or country "will get counted once" [9]. Multi-affiliation "further boosts inflation as now many authors list multiple institutions/countries in their affiliation" [9]. This inflation "casts doubts on the validity and reliability of university rankings and research evaluations employing bibliometric measures" [9]. If multiple affiliations are not accounted for, "each document counts as often as there are distinct affiliations reported on the publication" [7, p. 1041] which easily leads to distortions. The reliance of research rankings on bibliometric data can "create incentives for institutions to offer affiliations to researchers primarily employed elsewhere" [7, p. 1041]. The "most profound suspicion toward multi-affiliation is whether researchers and research institutions manipulate the system for immoral gains" [9]. A specific problematic practice is "paid affiliation" [3, p.63] which "involves paying individuals to list an institution as their affiliation in published research, even if they have no substantial connection to the institution" [3, p.64]. A "sharp and rapid rise in multi-affiliations within a short period" [3, p.82] especially across countries, may suggest "strategic efforts to amplify research output" [3, p.83]. To identify potentially questionable authorship and affiliation practices, two complementary "key metrics" [3, p.67] are suggested: research output growth and first authorship decline. A "marked reduction in first authorship may indicate shifts towards authorship and affiliation practices that could obscure genuine academic leadership" [3, p.67].

Despite their growing prevalence and the resulting concerns, multiple affiliations have "so far received little attention in the literature" [6, p. 382]. The phenomenon is generally "largely unexplored" [7, p. 1039]. This lack of systematic assessment is primarily due to the historical "lack of sufficient bibliometric data which would allow for determining such links" [2, p.286]. Bibliographic records previously failed to establish links between author affiliations and author names [4]. This constraint eased since 2008, when "a new author-institution tag available in Web of Science (WoS)" [2, p. 287] allowed researchers "to unambiguously link authors to their institutions and to determine whether an author has multiple affiliations" [2, p. 287]. Although a few previous studies have bibliometrically examined the multi-affiliation phenomenon in science publishing [9], "few related empirical studies have been conducted" [4, p. 557]. Consequently, systematic methodologies are still being developed "to identify researchers with multiple affiliations from bibliometric data" [1]. Given the scant systematic research, "we know very little about the phenomenon of researchers with multiple affiliations" [1]. This study introduces therefore a network-based framework to analyze multiple simultaneous institutional affiliations and their effects on institutional connectivity and citation impact. Using a Scopus data snapshot including papers published between 2013 and 2023 in Austria, the study distinguishes transient from persistent co-affiliations through parallel complete and temporally filtered networks. Combining gravity-type models with field- and year-normalized, fractionally counted citation indicators, the results demonstrate how geography and organizational type can shape co-affiliation patterns.

The paper is organised as follows. Section 2 introduces the data and outlines the preprocessing steps, describes the methods for estimating intra- and inter-institutional co-affiliation patterns, and details the procedures for assessing institutional and organisational citation impact. A subsequent subsection addresses potential threats to validity in the co-affiliation analyses. Section 3 presents and discusses the results, and Section 4 summarises the key findings and draws conclusions. All codes used for data processing, network construction and statistical modelling, as well as the curated datasets underlying the analyses, are openly available in public repositories (see Sections Code availability and Data availability).

## 2 Methods

### 2.1 Data

We conducted a Multiple Organisational Co-Affiliation Analysis (CAA) using the Scopus (Elsevier) bibliometric database [13]. All publications from 2013 to 2023 that included at least one author affiliated with an Austrian institution were retrieved, together with complete affiliation metadata. Following Yegros-Yegros et al. [1], publications in which at least one author reported two or more concurrent institutional affiliations within Austria were identified then.

The raw Scopus affiliation records required extensive disambiguation because of duplicate entries in different languages, incomplete addresses, address variants and heterogeneous representations of subunits (for example, departments or institutes within universities). To address these issues, all Austrian affiliations were harmonised at the parent-organisation level. Subunits were assigned to their corresponding parent organisations. For example, an affiliation reported as the Institute of Physics, Vienna University of Technology (TU Wien) was mapped to the parent organisation TU Wien. The same principle was applied across all eight organisation types considered in this study (Table 1):

universities (uni), research institutes (res), medical institutions (med), colleges (coll), companies (comp), government organisations (gov) and non-profit organisations (npo).

During harmonisation, Scopus affiliation identifiers (IDs) were mapped to unique organisation identifiers and duplicate entities were consolidated. Co-affiliations involving multiple subunits of the same parent organisation were removed, ensuring that the analysis focused on concurrent affiliations between distinct organisations, consistent with the approach of Yegros-Yegros et al. [1]. The entire dataset was independently harmonised by two researchers, after which any discrepancies were jointly reviewed and resolved to ensure accuracy and consistency. The harmonisation process resulted in 1647 unique Austrian organisations, each assigned to one of the ten organisational categories (Table 1). From harmonised affiliation and publication metadata, we constructed a network of inter-organisational co-affiliation links, defined as pairs of organisations concurrently listed by the same author. Authors reporting more than two simultaneous affiliations contributed all unique pairwise combinations (for example, an author affiliated with [A, B, C] generates the links A–B, A–C and B–C). We refer to this complete dataset of links as CoAffAll (Table 2).

Table 1: Organisation type classification of affiliations

| Organisation type | Description | Number of affiliations |
|---|---|---|
| uni (universities) | Universities, university medical centres, and university research collaborations. *Examples: University of Vienna; Medical University of Graz.* | 67 |
| res (research institutes) | Organisations conducting basic or applied research across diverse disciplines, including scientific centres, research societies and applied research institutes. Companies exclusively engaged in research activities are also included. *Examples: Austrian Academy of Sciences (ÖAW); AIT Austrian Institute of Technology.* | 225 |
| med (medical institutions) | Medical institutions providing care, treatment, and diagnostic services, including public and private hospitals, specialized clinics, and private medical practices. *Examples: Ottakring clinic; Salzburg Trauma Hospital.* | 251 |
| coll (colleges) | Universities of applied sciences (in German: Fachhochschulen), higher technical collages (HTL), etc. *Examples: University of Applied Sciences (FH) Campus Vienna; CAMPUS 02 University of Applied Sciences* | 29 |
| comp (companies) | Private companies, firms and consultancies. *Examples: Siemens Austria; AVL List GmbH.* | 552 |
| gov (government organisations) | Public bodies responsible for governance, administration, and policy implementation, including ministries, agencies, and public authorities. *Examples: Federal Ministry of the Republic of Austria for Social Affairs Health Care and Consumer Protection; Environment Agency Austria* | 98 |
| npo (non-profit-organisations) | Associations, foundations, and societies pursuing social, cultural, or public-interest goals, such as advocacy, humanitarian, or membership-based organizations. *Examples: Red Cross branches; Green Circle Society.* | 182 |
| other | Organisations not fitting into any of the categories above. | 243 |

To identify stable co-affiliations, we applied a temporal filter that retained only those organisational pairs that reappeared in publications by the same author in at least two distinct years separated by ≥2 years. This filtered dataset is termed CoAffStable (Table 2). For instance, if an author initially lists [A, B, C] and, two or more years later, lists [A, B], only the A–B link is preserved, whereas A–C and B–C are removed. We imposed a minimum gap of two years to reduce the inclusion of transient affiliations arising from short-term visits, sabbaticals, or brief employment transitions, which do not represent durable organisational relationships. Although this filter may omit some legitimate but short-lived co-affiliations, it ensures that the resulting network reflects enduring organisational ties [1,2,5]. A similar approach was used by Yegros-Yegros *et al.* [1], who required authors to report the same combination of affiliations in both 2016 and 2018. By insisting that at least one publication in each year list identical affiliations, their two-year filter ensures that only multi-year, simultaneous appointments were retained, removing entire author–year observations that did not meet this criterion. Our procedure, as illustrated above, operates at the link level, retaining the recurring A–B link while discarding non-persistent pairs. Rather than excluding an author or their publications outright, we removed only those specific affiliation pairs that failed to recur across the required temporal window of two years. This yields a more granular filter that preserves all other stable links associated with the same author.

The Scopus-derived bibliographic article dataset and the processed affiliation metadata generated through the harmonisation procedure are publicly available; details and persistent identifiers are provided in the Data availability Section.

Table 2. Summary of co-affiliation link datasets used in the analysis

| Dataset Name | Description | Criteria | Number of links |
| --- | --- | --- | --- |
| CoAffAll | Contains all pairwise links between distinct organisations for authors with multiple concurrent affiliations. | Includes all unique pairwise combinations of affiliations for authors with two or more affiliations. | 44298 |
| CoAffStable | Subset of the above dataset restricted to stable co-affiliations. | Includes only links where authors maintain the same pair of affiliations in at least two distinct years | 29881 |

## 2.2 Organisational proximity

To quantify the intensity of collaboration between affiliations, we employed a gravity model which is widely used in economics and regional science. We employed the model to explain interactions among entities, such as regions or organisations, as a function of their attributes (for example, size, research output or funding) and the distance between them. Although inspired by Newton's law of universal gravitation, in which physical mass and spatial distance govern attraction, the economic analogue replaces physical mass with organisational characteristics. The economic analogue interprets distance primarily as geographical separation, while also allowing extensions to non-spatial forms of separation (for example, institutional or cultural) [14,15]. In this setting, the gravity model provides a quantitative basis for assessing how organisational features (see Table 1) and organisational proximity influence the strength of collaborative ties between affiliations [1,16,17].

The basic gravity model is formulated as follows:

$$I_{ij} = G \frac{M_i^\alpha M_j^\beta}{D_{ij}^\gamma} \#(1)$$

where $I_{ij}$ represents the interaction intensity between two masses $i$ and $j$; $G$ is a proportionality constant; $M_i$ and $M_j$ denote the masses of the entities; and $D_{ij}$ is the distance between them.

To facilitate empirical estimation, the model is typically expressed in a log-linear form, allowing parameter estimation via linear regression techniques. The linear regression can be written as:

$$lnI_{ij} = lnG + \alpha_1 lnM_i + \alpha_2 lnM_j - \beta lnD_{ij} \#(2)$$

In our study, the interaction $I_{ij}$ represents an undirected interaction derived from co-affiliation links. Therefore, distinguishing between $M_i$ and $M_j$ is unnecessary, and we set $\alpha_1 = \alpha_2 = \alpha$ [1,18]. Consequently, the regression model can be simplified to:

$$lnI_{ij} = lnG + \alpha ln(M_i M_j) - \beta lnD_{ij} \#(3)$$

As the dependent variable in our study (interaction intensity between affiliation pairs) is count-based, we followed previous studies [1,14,15,17] and employed a non-linear specification. Specifically, we used a Poisson regression model estimated via maximum likelihood, in which the interaction intensity between two affiliated entities (*i* and *j*) follows a Poisson distribution with a conditional mean (μ). The distribution and the mean are a function of the explanatory variables.

$$Pr[I_{ij}] = \frac{e^{-\mu_{ij}} \mu_{ij}^{I_{ij}}}{I_{ij}!}, where\ \mu_{ij} = exp(G + \alpha ln(M_i M_j) - \beta ln D_{ij}) \#(4)$$

We propose two gravity models to examine how different forms of organisational proximity shape collaboration patterns: an intra-organisational proximity model (Model 1) and an inter-organisational proximity model (Model 2). Model 1 captures within-type proximity, in which both affiliations in a co-authorship pair belong to the same organisational category (for example, university–university or research-institute–research-institute links). Model 2 captures between-type proximity, where the two affiliations originate from different organisational categories (for example, university–research-institute or university–industry links). This distinction aligns with the modelling approach adopted by Yegros-Yegros et al. [1]. Both of our models extend the classical gravity-model framework by incorporating organisational proximity measures tailored to these distinct organisational relationships. The variables included in each specification are summarised in Table 3.

To construct the model inputs, we developed two undirected affiliation networks based on unique affiliations identified in the CoAff_All and CoAff_Stable datasets (Table 2). These networks include the primary organization types—universities (uni), research institutes (res), medical institutions (med), companies (comp), colleges (coll), non-governmental organizations (npo), and government institutions (gov). Each node in the network represents a unique affiliation, while edges denote connections between affiliations. Edge weights correspond to the number of co-linkages between paired affiliations (Table 3; affiliation_edge_strength). Node weights equal the sum of all connected edge weights, representing the total number of articles associated with each affiliation (Table 3; article_count_affiliation_from, article_count_affiliation_to). The total number of edges in the two networks corresponds to 1386 × (1386 - 1)/2 and 439 × (431 - 1)/2 unique organizational pairs, respectively.

Travel times between affiliation pairs were estimated by first geocoding all addresses to geographic coordinates in the World Geodetic System 1984 (WGS 84) reference frame using the Python geopy library. A local instance of the open-source routing engine Valhalla was then employed to compute travel times between each pair of affiliations based on these coordinates. Valhalla uses OpenStreetMap data to generate realistic travel routes and durations [19]. In classical gravity models, geographic distance typically serves as the impedance term representing spatial separation. Here, we replace distance with travel time, as it offers a more realistic measure of effective connectivity.

Unlike straight-line distance which assumes homogeneous transport conditions, travel time incorporates variations in infrastructure quality, routing constraints and border delays, thereby capturing more meaningful frictions that shape spatial interactions [17,20].

In the networks constructed from the CoAff–All and CoAff–Stable datasets, the variable affiliation_edge_strength contains a large proportion of zero values, as most organisational pairs do not share researchers who generate co-affiliation links. To address both overdispersion and the prevalence of zero counts, we employed a zero-inflated negative binomial (ZINB) regression model. This specification accounts for structural zeros (cases where the absence of links arises from a distinct underlying process rather than random variation) and is widely used in gravity-model applications, including trade analyses. In these analyses, zero flows between entity pairs are common [14]. The ZINB model consists of two components: a logit component that estimates the probability of a structural zero, and a negative binomial component that models the count of interactions for non-zero observations [14,17].

Table 3. Variables included in the two zero-inflated negative binomial (ZINB) regression models: the intra-organisational proximity model (Model 1) and the inter-organisational proximity model (Model 2)

| Variable | Description |
| --- | --- |
| Gravity model variables (applied in Model 1 and Model 2) | |
| affiliation_edge_strength ($I_{ij}$) [1/1] | The number of links between unique affiliation pairs (Intensity $I_{ij}$ between organisations). |
| article_count_affiliation_from ($M_i$) [1/1] | Total number of articles associated with the first affiliation (from_node) in a pair (Mass $M_i$). |
| article_count_affiliation_to ($M_j$) [1/1] | Total number of articles associated with the second affiliation (from_node) in a pair (Mass $M_j$). |
| travel_time ($D_{ij}$) [s] | Time required in seconds, to travel from one organization to the other (corresponds to the gravity model distance in $D_{ij}$ between the affiliations pairs). |
| Proximity variables | |
| Intra-institutional proximity (Model 1) [1/1] | Seven dummy variables indicating whether both affiliations belong to the same organisational category (1 = same type; 0 = otherwise): uni_uni, res_res, med_med, comp_comp, coll_coll, ngo_npo, gov_gov. |
| Inter-institutional proximity (Model 2) [1/1] | Twenty-one dummy variables indicating whether affiliations belong to different organisational categories (1 = different types; 0 = otherwise): coll_comp, coll_gov, coll_med, coll_npo, coll_res, coll_uni, comp_gov, comp_med, comp_npo, comp_res, comp_uni, gov_med, gov_npo, gov_res, gov_uni, med_npo, med_res, med_uni, ngo_res, ngo_uni, res_uni. |

## 2.3 Citation impact analyses

To quantify citation impact, we employed field- and year-normalised citation scores based on Hazen percentiles, following the methodological recommendations of Bornmann et al. [21]. For each publication, we computed a median weighted Hazen percentile rank, which serves as a normalised indicator of citation impact. The indicator enables direct comparisons across research fields and publication years by accounting for disciplinary and temporal citation differences. Fields are defined as journal sets in Scopus. The aggregated percentile rank is defined as a weighted percentile, with weights given by the number of papers in each field × publication year reference set. This weighting accounts for differences in both field-specific citation practices and publication-year citation windows, ensuring that citation percentile information derived from larger and temporally comparable paper sets contributes proportionally more to the aggregated indicator than information from smaller sets. The resulting weighted percentile rank (wPR) therefore represents a field- and year-normalised citation score for publications assigned to multiple subject categories. The wPR was calculated as [21]

$$\text{wPR} = \frac{\sum_x \text{PR}_{SC_x}\, n_{SC_x}}{\sum_x n_{SC_x}} \#(5)$$

where $\text{PR}_{SC_x}$ denotes the Hazen percentile of the paper in the field $SC_x$ for the corresponding publication year, and $n_{SC_x}$ is the total number of papers in that field and publication year.

Because individual researchers may hold multiple organisational affiliations simultaneously and because each organisation is classified into an organisational type in this study, citation impact was allocated in a way that treats these overlapping memberships consistently. For example, a researcher may be affiliated with both TU Wien (classified as a university) and the Austrian Academy of Sciences (classified as a research institute). When evaluating organisational type performance and individual organisations, each publication was allocated thus proportionally to all associated affiliations, either across organisational types (e.g., 0.5 to a university and 0.5 to a research institute) or, at the organisation level, across the corresponding organisations (e.g., 0.5 to the TU Wien and 0.5 to the Austrian Academy of Sciences). To ensure equitable allocation of citation impact across all units to which multi-affiliated authors belong, wPR values were aggregated using fractional counting. This approach guarantees that each publication contributes only a proportional share of its citation impact to every organisation and to every organisational type. For any given unit $F$ we computed the mean weighted percentile rank, $\text{mwPR}(F)$. It is the fractional weighted mean of the wPR values belonging to its constituent publications

$$mwPR(F) = \frac{\sum_{i=1}^{y} wPR_i \cdot FR_i}{\sum_{i=1}^{y} FR_i} \#(6)$$

where $wPR_i$ denotes the median weighted Hazen percentile rank of paper *i*, $FR_i$ denotes the fractional contribution of paper *i* to unit *F*, and *y* denotes the total number of papers associated with *F* [22,23].

To assess the sensitivity of citation-impact estimates to alternative representations of author affiliations, the aggregation was repeated under three affiliation-filtering schemes (Table 4). In the All-Affiliations (AA) scheme, every affiliation and its corresponding organisational type from each inter-organisational link was included. The First-Affiliation (FA) scheme retained only the first-listed affiliation from each link, whereas the Last-Affiliation (LA) scheme retained only the last-listed one. Using the link structure described above (see the example in Section 2.1 illustrating links generated from author affiliations [A, B, C]), these schemes correspond to (1) including all affiliation links, (2) keeping only the first-listed affiliations (e.g., A from each pair), or (3) keeping only the last-listed affiliations (e.g., C). This sensitivity analysis addresses the fact that multi-affiliated authors can redistribute citation credit across organisations depending on how their affiliations are represented. For instance, a researcher jointly affiliated with a university and a research institute may inflate the apparent performance of one unit over the other if only a single affiliation is retained. More broadly, organisational rankings may shift due to seemingly arbitrary choices in affiliation filtering. By comparing results across the AA, FA and LA schemes, we assessed the robustness of citation-impact estimates to these alternative representations of multi-affiliated authors and quantified the extent to which such modelling decisions influenced the allocation of citation credit across organisation types and individual organisations.

To evaluate how sensitive institutional rankings are to the volume of publications attributed to each organisation, we conducted an additional sensitivity analysis based on minimum sample-size thresholds. Organisational mwPR values were recalculated under a series of publication-count cutoffs applied to the entire analysis period (2013–2023). Only organisations whose publication volume exceeded the specified threshold were included in each iteration. This procedure assesses how stable organisational rankings remain when excluding small organisations whose citation-impact estimates are highly volatile due to low publication counts. By comparing mwPR-based rankings across increasing thresholds, we quantified the extent to which organisational performance assessments depend on minimum publication requirements. We ensured thus that subsequent comparisons focussed on units with sufficiently robust empirical support. This minimum sample-size requirement limited distortions that arise when organisations with very small publication portfolios are ranked alongside substantially larger organisations. Organisations with only a few publications can appear disproportionately strong if a single highly cited output yields an exceptionally high wPR value, whereas larger organisations—such as major universities—typically exhibit a broad distribution of citation performance, resulting in more stable but often more moderate mwPR estimates.

Table 4. Affiliation-filtering schemes. Each scheme specifies which affiliations from inter-organisational links are retained when aggregating citation data. For example, if an author lists three concurrent affiliations (A–B–C), the link structure yields three inter-organisational pairs: (A–B), (A–C) and (B–C). Under the All-Affiliations (AA) scheme, all links are retained. The First-Affiliation (FA) scheme keeps only the first-listed affiliations from each pair (A, A), whereas the Last-Affiliation (LA) scheme keeps only the last-listed affiliations (C, C).

| Schema | Abbreviation | Description | Example outcome |
| --- | --- | --- | --- |
| All affiliations | AA | All affiliation links included | (A – B), (A – C), (B – C) |
| First-affiliation filter | FA | Only first-listed affiliation retained. | (A), (A) |
| Last-affiliation filter | LA | Only last-listed affiliation retained | (C), (C) |

## 2.4 Threats to validity

A key source of potential bias arises from the classification of institutions into organisational types. This taxonomy is, to some extent, subjective and can influence both the ZINB modelling results and the mwPR-based performance estimates. For example, businesses with substantial research activity were classified as research institutes, though they could alternatively be categorised as companies. Similarly, university medical centres were assigned to the university category, even though they might also be considered as medical institutions. Different plausible classification choices would yield different aggregations of publications and, consequently, different impact estimates. The organisational-type schema should therefore be interpreted as an analytical choice aligned with the evaluative focus of the study, rather than as a definitive representation of institutional boundaries.

The difficulty of delineating organisational types is well illustrated by the methodology of the Centre for Science and Technology Studies (CWTS) in producing the Leiden Ranking. CWTS notes that no internationally accepted criteria exist for defining what constitutes a "university," and that national academic systems differ markedly in how entities such as university medical centres, research institutes and affiliated hospitals are structured. As a result, CWTS must make normative judgments about whether such organisations should be treated as components of a university, joint facilities or independent entities, decisions that directly affect how publications are attributed and how institutions are compared [24,25]. These challenges are even more pronounced in our study, which spans multiple organisational types, including research institutes, medical institutions, companies and non-profit organisations. Each category contains ambiguous cases that resist clear-cut classifications.

Another source of uncertainty arises from organisational sample sizes. As described above, the stability of organisational mwPR estimates depends on the number of publications attributed to each organisation. Organisations with very small output volumes can exhibit highly volatile citation-impact values, which may distort comparative rankings. Although we mitigated this issue by applying minimum publication thresholds and by examining how rankings changed across increasingly stringent cutoffs, the choice of threshold itself introduced a modelling decision that may influence the results.

# 3 Results

## 3.1 Analyses of organisational proximity

In line with prior findings reported in the introduction, our analysis shows that 21,042 out of 83,960 articles (25%) include at least one author with multiple affiliations, closely matching the previously reported average that "about one quarter of articles" exhibit multiple affiliations [7]. Figure 1 shows two network graphs derived from the datasets summarised in Table 1, displaying only nodes and edges between organisations with an edge strength of 250 or greater. Major organisations with extensive research activity appear in both the complete (CoAff–All, Figure 1a) and filtered networks (CoAff–Stable, Figure 1b). In the filtered CoAffStable version, several smaller nodes and weaker connections are removed, producing a more consolidated overall structure. In both networks, universities and research institutes dominate. The largest nodes, where size corresponds to the total number of publications per organisation, are primarily universities (Figure 1, purple) followed by research institutes (green). Universities are connected predominantly with other universities and research institutes. In both datasets, strong co-affiliation links occur between the Austrian Academy of Sciences (ÖAW Wien), the Technical University of Vienna (TU Wien) and the University of Innsbruck (Uni Innsbruck), indicating a high concentration of researchers with simultaneous affiliations at these organisations A smaller cluster represents hospital organisations, while individual, isolated nodes correspond to other organisational types such as companies, non-profit organisations, and government bodies.

Table 5 summarises the results of the ZINB regression for Model 1, applied to both the CoAff–All and CoAff–Stable link datasets. This intra-organisational proximity model assesses how co-affiliations within the same organisational sphere, together with spatial distance, shape the likelihood of collaborative link formation between authors. Across both datasets, greater travel time is associated with a reduced probability of co-affiliation links, highlighting the enduring role of geographical proximity in shaping patterns of simultaneous affiliations. The variable ln_prod_edge_strength—the logarithm of the product of the two organisation's publication counts—is positively and significantly associated with link formation in both datasets. This means that organisations with larger overall publication volumes are more likely to be connected through co-affiliations. The pattern thus reflects a size-dependent mechanism: larger institutions have more researchers, and therefore a higher probability that some of them maintain multiple concurrent affiliations. Crucially, in our setting this effect reflects organisational publication capacity and the likelihood of hosting multi-affiliated researchers. It does not capture collaborative activity or joint productivity between organisations.

Organisational pairing effects further illuminate structural patterns: university–university (univ_univ) links exhibit strong and statistically significant positive coefficients in both datasets, underscoring universities' central role as hubs within the organisational collaboration landscape. Positive and statistically significant coefficients for medical–medical (med_med) pairings in the CoAffAll dataset indicate that authors frequently hold multiple concurrent affiliations within medical institutions. In the CoAffStable dataset, however, these effects weaken or lose statistical significance. This attenuation likely reflects the stricter inclusion criteria. Short-term co-affiliations with corresponding low publication activities were excluded—such as clinical fellowships or adjunct teaching roles (many authors appeared only once). The coefficient for coll_coll is negative and statistically insignificant in the CoAffAll model but turns positive (and remains statistically insignificant) in the CoAffStable model, suggesting no robust or consistent effect of collaboration–collaboration pairings on co-affiliation link formation.

Company–company (comp_comp) links show negative coefficients in the ZNIB model, particularly in the CoAffStable dataset, indicating that co-affiliations between two private-sector organisations are comparatively uncommon. This pattern may reflect structural features of the private sector, such as contractual restrictions, confidentiality requirements, and limited incentives for shared appointments, which reduce the likelihood that researchers formally hold concurrent affiliations across companies. Although non-profit (npo_npo) and government–government (gov_gov) pairings are statistically not significant in the full dataset, the npo_npo coefficient becomes strongly positive in the filtered CoAffStable results (but with a very large standard error). This high imprecision indicates that the apparent positive effect is not statistically robust and provides no reliable evidence for strengthened or durable NPO co-affiliations. The shift suggests that, although NPOs are infrequently co-affiliated overall, the filtered dataset does not support the presence of a particularly stable or cohesive NPO sub-network.

In the inflation component, the coefficient on joint productivity is negative and statistically significant, indicating that higher combined productivity of author pairs is associated with a lower likelihood of structural zeros (that is, fewer pairs with no latent potential to form co-affiliation ties between organisations). At the same time, the positive and significant intercept points to a substantial baseline probability of excess zeros, consistent with the sparse and zero-inflated structure of organisational co-affiliation networks.

(a) CoAffAll

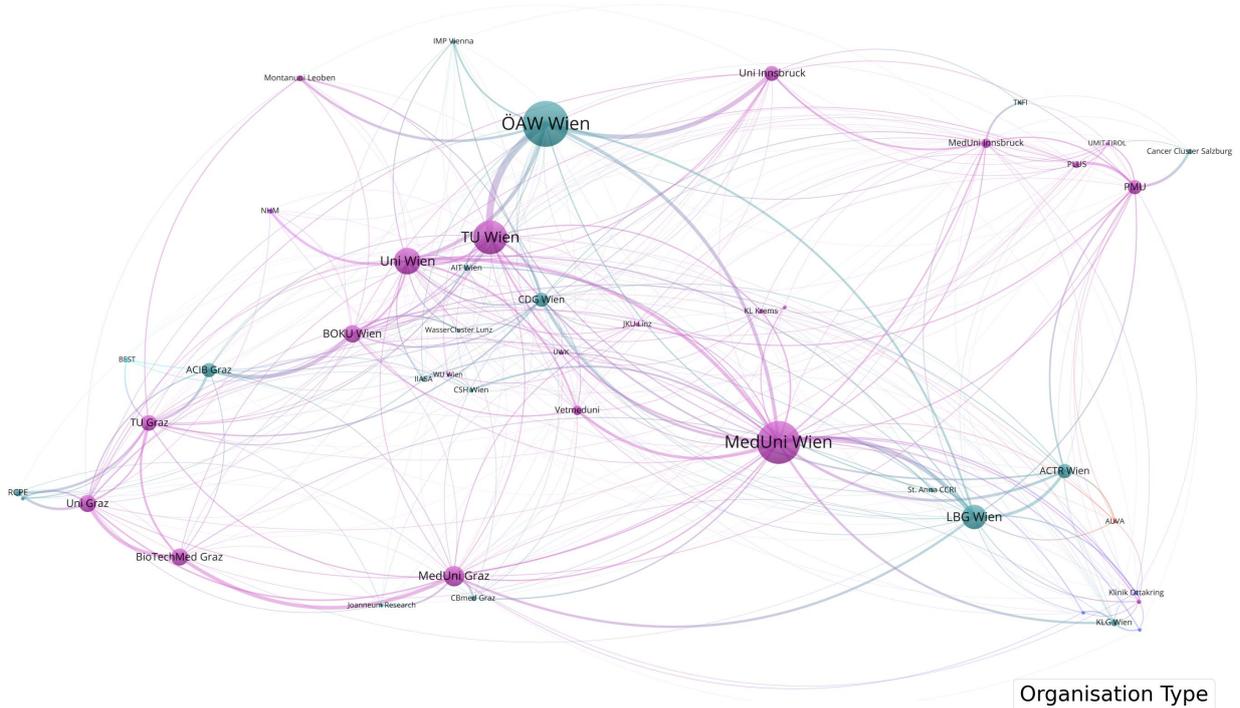

(b) CoAffStable

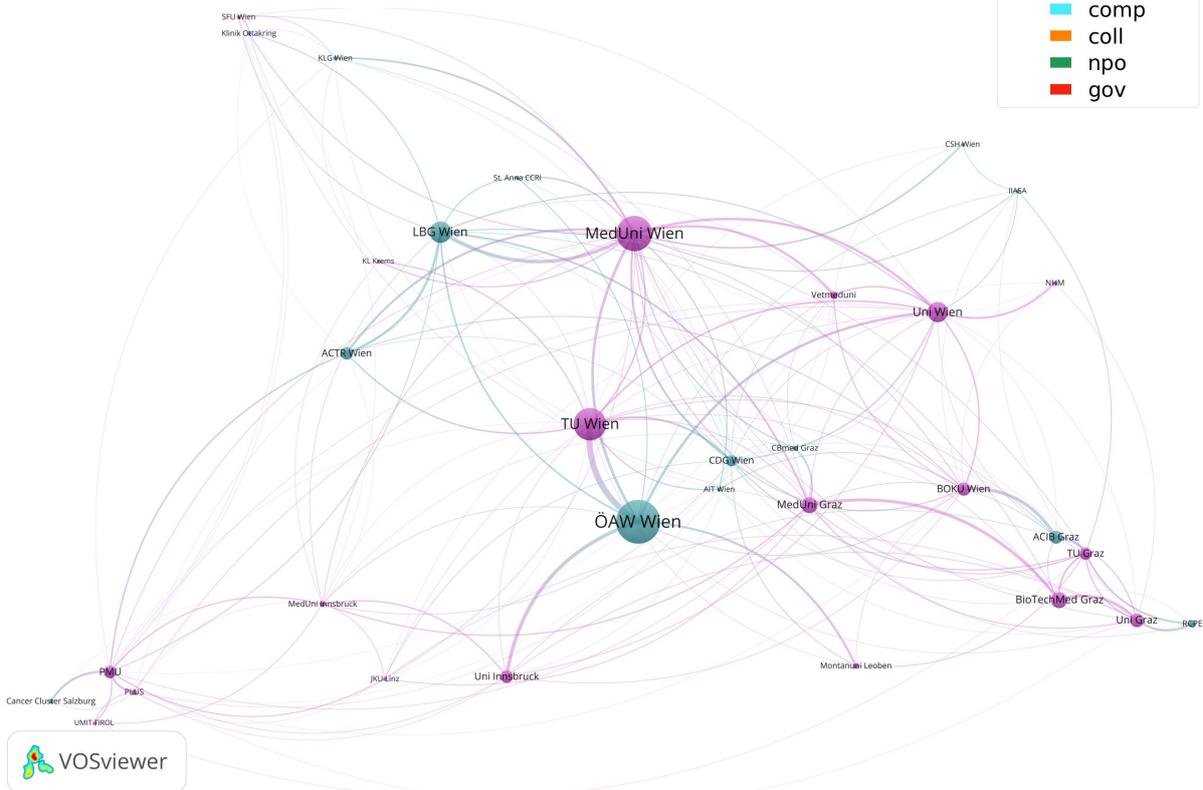

Figure 1. Co-affiliation networks based on the complete (a) and filtered (b) datasets. Nodes represent organizations (sized by publication output; coloured by organization type), and edges indicate co-affiliation links with strength ≥ 250.

Organisations: universities (uni), research institutes (res), medical institutions (med), colleges (coll), companies (comp), government organisations (gov), and non-profit organisations (npo).

Table 5. Zero-inflated negative binomial regression results for the intra-organisational proximity Model 1

| Results Model 1 | CoAffAll | | | CoAffStable | | |
|---|---|---|---|---|---|---|
| | Coef. | Robust Std. err. | P>\|z\| | Coef. | Robust Std. err. | P>\|z\| |
| ln_prod_edge_strength; $ln(M_i M_j)$ | 0.25 | 0.02 | 0.00 *** | 0.23 | 0.03 | 0.00 *** |
| ln_travel_time; $ln\,D_{ij}$ | -0.56 | 0.05 | 0.00 *** | -0.60 | 0.08 | 0.00 *** |
| uni_uni | 2.40 | 0.16 | 0.00 *** | 1.64 | 0.20 | 0.00 *** |
| res_res | -0.11 | 0.25 | 0.65 | -0.69 | 0.32 | 0.03 * |
| med_med | 0.33 | 0.18 | 0.06 . | -0.15 | 0.38 | 0.69 |
| comp_comp | -0.54 | 0.18 | 0.00 ** | -1.46 | 0.46 | 0.00 ** |
| coll_coll | -0.31 | 0.61 | 0.62 | 0.71 | 0.87 | 0.41 |
| npo_npo | 0.63 | 0.38 | 0.10 | 2.85 | 0.95 | 0.00 ** |
| gov_gov | -0.27 | 1.07 | 0.80 | -1.57 | 0.81 | 0.05 . |
| Logit Model | | | | | | |
| inflate_const | 5.58 | 0.16 | 0.00 *** | 6.03 | 0.24 | 0.00 *** |
| inflate_ln_prod_edge_strength | -0.33 | 0.01 | 0.00 *** | -0.36 | 0.02 | 0.00 *** |
| constant | 1.03 | 0.57 | 0.07 . | 2.85 | 0.78 | 0.00 *** |
| N | 959805 | | | 92665 | | |
| BIC | 43849.56 | | | 14435.52 | | |
| AIC | 44002.63 | | | 14558.19 | | |

Notes. Reported are coefficients (Coef.), robust standard errors (Std. err.), and p-values. Significance levels: *p < 0.05, **p < 0.01, ***p < 0.001. AIC = Akaike Information Criterion. BIC = Bayesian Information Criterion.

The results of Model 2 are shown in Table 6 for both datasets. Two robust patterns dominate the count component, that is the negative binomial part of the model. First, the distance-decay effect from Model 1 persists: greater travel time is associated with substantially lower expected counts of co-affiliation links (ln_travel_time = −0.82, $p < 0.001$; −0.80, $p < 0.001$). Second, joint publication strength is a consistent predictor of co-affiliation formation. However, its influence attenuates with temporal filtering: the ln(prod_edge_strength) coefficient is positive in the full CoAffAll network (0.62, $p < 0.001$) and remains positive (albeit substantially reduced) in the CoAffStable sample (0.3, $p < 0.01$). Together these patterns suggest that joint productivity contributes to link formation overall (for example, via project-level or short-term collaborations) but explains less of which inter-organisational ties endure over multiple years.

Organisational pairings reveal a clear hierarchy. University–research-institute links are uniquely resilient. They are the only cross-type pairing with a positive and statistically significant coefficient in both samples (resi_uni = 0.98, $p < 0.001$; 0.78, $p < 0.05$). By contrast, many cross-type combinations (particularly those involving government, non-profit and medical institutions) carry large negative coefficients, and several of these become more negative after filtering (for example, coll_med: −1.80 → −8.92; comp_npo: −3.11 → −6.66; coll_res: −0.31 → −1.55). Despite these shifts, the pattern is not universal: some pairings weaken or lose statistical significance in the filtered network (e.g., coll_uni and gov_med), and several estimates remain statistically non-significant (notably coll_comp: -0.70, $p$ = 0.60 → −0.61, $p$ = 0.48). Overall, temporal filtering reduces the footprint of short-lived, project-based affiliations. Temporal filtering also sharpens sectoral distinctions, leaving a smaller set of organisational relationships that are likely to reflect enduring organisational linkages or formal partnerships.

Finally, the inflation component confirms that higher joint productivity reduces the probability of structural zeros (inflate_ln_prod_edge_strength = −1.30, −1.42; both $p < 0.001$), while the positive intercept reflects the overall sparsity of inter-organisational co-affiliations.

Table 6. Zero-inflated negative binomial regression results for the inter-organisational proximity model (Model 2), analogous to the intra-organisational model (Model 1) presented in Table 5 but incorporating cross-type organisational pairings

| Results Model 2 | CoAffAll | | | CoAffStable | | |
|---|---|---|---|---|---|---|
| | Coef. | Robust Std. err. | P>\|z\| | Coef. | Robust Std. err. | P>\|z\| |
| ln_prod_edge_strength; $ln(M_i M_j)$ | 0.62 | 0.06 | 0.00 *** | 0.33 | 0.10 | 0.00 ** |
| ln_travel_time; $ln\ D_{ij}$ | -0.82 | 0.04 | 0.00 *** | -0.80 | 0.08 | 0.00 *** |
| coll_comp | -0.70 | 0.60 | 0.24 | -0.61 | 0.85 | 0.48 |
| coll_gov | -2.82 | 0.59 | 0.00 *** | -25.04 | 73.91 | 0.73 |
| coll_med | -1.80 | 0.22 | 0.00 *** | -8.92 | 0.19 | 0.00 *** |
| coll_npo | -0.85 | 0.39 | 0.03 * | -5.60 | 0.20 | 0.00 *** |
| coll_res | -0.31 | 0.32 | 0.34 | -1.55 | 0.57 | 0.01 ** |
| coll_uni | 1.18 | 0.27 | 0.00 *** | -0.29 | 0.37 | 0.43 |
| comp_gov | -2.95 | 0.45 | 0.00 *** | -3.75 | 0.95 | 0.00 *** |
| comp_med | -3.05 | 0.42 | 0.00 *** | -3.19 | 0.71 | 0.00 *** |
| comp_npo | -3.11 | 0.35 | 0.00 *** | -6.66 | 0.82 | 0.00 *** |
| comp_res | -2.43 | 0.19 | 0.00 *** | -2.82 | 0.33 | 0.00 *** |
| comp_uni | -1.65 | 0.16 | 0.00 *** | -1.76 | 0.22 | 0.00 *** |
| gov_med | -0.83 | 0.43 | 0.05 . | -0.77 | 0.74 | 0.30 |
| gov_npo | -1.93 | 0.49 | 0.00 *** | -2.78 | 0.92 | 0.00 ** |
| gov_res | -0.71 | 0.59 | 0.23 | -0.29 | 0.79 | 0.72 |
| gov_uni | -0.90 | 0.24 | 0.00 *** | -0.93 | 0.32 | 0.00 ** |
| med_npo | -0.78 | 0.21 | 0.00 *** | -1.30 | 0.45 | 0.00 ** |
| med_res | -1.33 | 0.22 | 0.00 *** | -1.86 | 0.33 | 0.00 *** |
| med_uni | -0.70 | 0.16 | 0.00 *** | -0.94 | 0.19 | 0.00 *** |
| npo_res | -2.24 | 0.49 | 0.00 *** | -2.98 | 0.80 | 0.00 *** |
| npo_uni | -1.46 | 0.24 | 0.00 *** | -1.22 | 0.39 | 0.00 ** |
| resi_uni | 0.98 | 0.25 | 0.00 *** | 0.78 | 0.31 | 0.01 * |
| Logit Model | | | | | | |
| inflate_const | 4.67 | 0.12 | 0.00 *** | 4.28 | 0.20 | 0.00 *** |
| inflate_ln_prod_edge_strength | -1.30 | 0.04 | 0.00 *** | -1.42 | 0.06 | 0.00 *** |
| constant | -1.21 | 0.21 | 0.00 *** | 1.23 | 0.34 | 0.00 *** |
| N | 959805 | | | 92665 | | |
| BIC | 42887.26 | | | 14211.05 | | |
| AIC | 43205.17 | | | 14465.85 | | |

Notes. Reported are coefficients (Coef.), robust standard errors (Std. err.), and p-values. Significance levels: *$p < 0.05$, **$p < 0.01$, ***$p < 0.001$. AIC = Akaike Information Criterion. BIC = Bayesian Information Criterion.

## 3.2  Impact of co-affiliated research

Figure 2 illustrates the probability density of wPR for the different organizational types under the filtering schemes described in Section 3.2, using both datasets (CoAffAll and CoAffStable; see Table 2).

The overall distribution patterns within each organisational type remain consistent across the three schemes (all, first and last). Research institutes (res) and universities (uni) exhibit broader violin shapes at higher wPR values, indicating a greater concentration of publications with above-average, field- and year-normalized citation impact. In contrast, government organisations (gov), colleges (coll) and non-profit organisations (npo) display narrower distributions toward the upper wPR range, suggesting fewer highly cited publications. The companies (comp) and medical institutions (med) types show a more even spread of wPR values above 50% in both datasets. Overall, all organizational types exhibit a comparably wider distribution above the 50% line, indicating that most outputs perform at least around or above the expected value for an 'average' citation level.

The ordering of mwPRs (diamonds in Figure 2) across all organizational types is consistent between datasets CoAffAll and CoAffStable. For instance, research institutes (res) consistently follow the pattern last > all > first, whereas universities (uni) follows first > all > last in both datasets. The all-affiliations filter systematically yields intermediate values, reflecting the averaging of a multi-affiliated author's contribution across all listed affiliations. Across organisational types, research institutes (res) and universities (uni) demonstrate the highest overall citation impact. Research institutes (res) show their strongest performance when last-listed affiliations are considered, whereas universities (uni) perform best when first-listed affiliations are retained. Among the remaining types, the most notable variability across filtering schemes appears for government organisations (gov), where mwPR under the FA filter in the CoAff–Stable dataset is substantially lower than in CoAff–All.

These results show that mwPR estimates are sensitive to which multi-affiliated author's listed affiliations are used. Nevertheless, the relative ordering of the FA, AA, and LA schemes remains stable across datasets and organisational types, indicating a robust pattern in how affiliation position shapes observed citation impact.

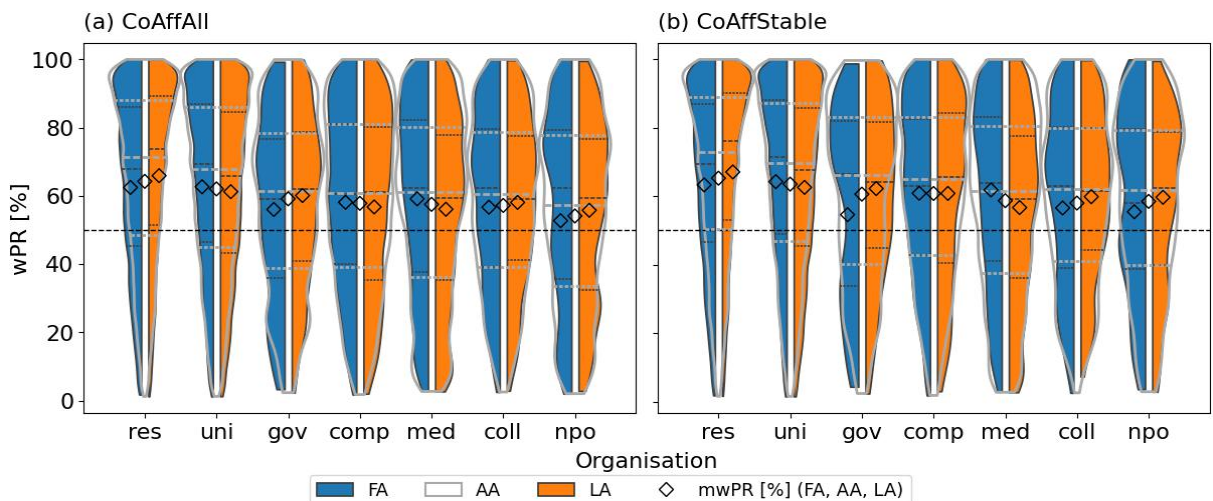

Figure 2: Distribution of median weighted Hazen percentile ranks (wPR) across organizational types (res, uni, gov, comp, med, coll, npo) for the datasets CoAffAlls (a) and CoAffStable (b). Violin plots show the probability density of wPR for each organizational type under the first- (FA), all- (AA), and last-affiliation (LA) filters. Diamonds indicate the corresponding mean weighted percentile ranks (mwPR). Table 2 for more information about the link datasets. Organisations: universities (uni), research institutes (res), medical institutions (med), colleges (coll), companies (comp), government organisations (gov), and non-profit organisations (npo).

We examined temporal trends in citation impact by aggregating mwPR annually for each organisational type in the CoAffAll dataset (Figure 3a) and CoAffStable dataset (Figure 3b). In 2013, research institutes (res) exhibit the highest mwPRs (Figure 3a: 68%; Figure 3b: 70%), with government organisations (gov) close behind (Figure 3a: 67%; Figure 3b: 69.5%). Research organizations consistently rank first, with mwPRs generally in the mid- to high-60s, while universities (uni), which account for the largest share of co-affiliation links, maintain steady, moderate performance around the low 60s. Government organisations (gov) and non-profit organisations (npo) have the greatest temporal variability: Government organisations (gov) decline from its 2013 peak (≈67% in Figure 3a; ~70% in Figure 3b) to values near the low 50s in later years (e.g., ≈50% in 2017 and ≈53% in 2021 in Figure 3a). Non-profit organisations (npo) fluctuate substantially across the series (e.g., ≈58% in 2013, ≈50% in 2014, and >64% in 2015 in Figure 3a). By contrast, companies (comp) vary more modestly and remain comparatively stable across both datasets. Medical institutions (med) and colleges (coll) occupy the lower mwPR range (mid- to high-50s) showing small fluctuations and a mild rebound in the most recent year.

Across nearly all organisational types, the CoAffStable dataset yields systematically higher mwPR values than the unfiltered dataset despite smaller sample sizes, indicating that persistent co-affiliations are associated with higher citation impact. These temporal patterns align with the distributional differences illustrated in Figure 2 and underscore both the heterogeneity of citation performance across organisational types and the sensitivity of impact estimates to how co-affiliations are defined.

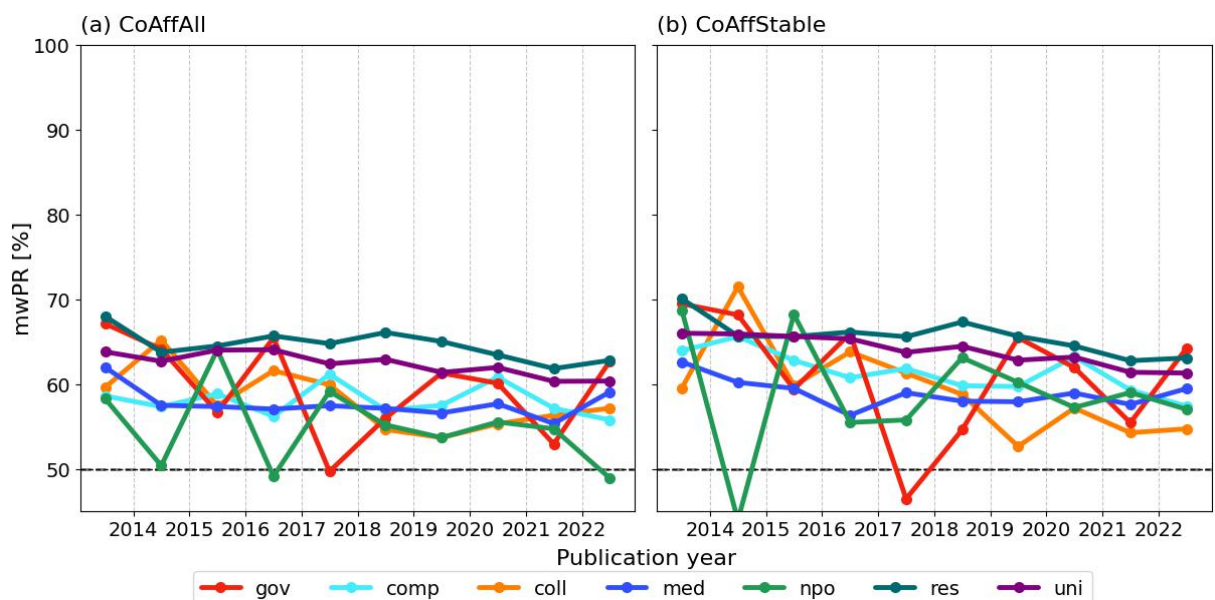

Figure 3: Temporal evolution of citation impact across organisational types. Annual mean weighted percentile ranks (mwPR) by organisational type for (a) affiliations_links_all and (b) CoAffStable. Values are aggregated per year. See Table 2 for more information about the link datasets. Organisations: universities (uni), research institutes (res), medical institutions (med), colleges (coll), companies (comp), government organisations (gov), and non-profit organisations (npo).

Figure 4 summarises mwPR values for the ten top-performing Austrian research organisations, identified based on their mwPR in the CoAff–All dataset. The figure includes only those organisations with at least 300 publications in the unfiltered CoAff–All dataset (AA scheme), representing an intermediate level of filtering. The resulting top-ten list of organisations is exclusively dominated by universities and research institutes. Results for alternative threshold levels, illustrating how the composition of the top-ranked organisations shift with different sampling requirements, are presented in the Appendix.

All organisations in Figure 4 fall into the universities (uni) or research institutes (res) categories. For these same organisations, mwPR values were also calculated using the CoAffStable dataset, enabling direct comparison between unfiltered and stable co-affiliation structures. Across both datasets, the Austrian Academy of Sciences (ÖAW) Vienna consistently leads, with mwPR values of 73% (n = 2,960) in the AA dataset and 76% in the filtered dataset (n = 2,559). The ÖAW also attains the highest scores under the LA scheme (CoAffAll LA: 76%, n = 1,808; CoAffStable LA: 77%, n = 1,530).

In the CoAffAll AA dataset, the International Institute for Applied Systems Analysis, Laxenburg (IIASA; 72.5%, n = 355) and TU Wien (65.1%, n = 2,550) follow the ÖAW.

Applying the FA and LA filters introduces systematic shifts across organisations. FA values are generally similar to or slightly lower than AA values for most organisations, though TU Wien exhibits a modest increase under FA. Several organisations show pronounced declines under LA (most notably TU Wien (56%, n = 680) and IIASA (65%, n = 74). When moving from the full dataset (CoAffAll) to the filtered dataset (CoAffStable), mwPR increases for nearly all organisations. This systematic increase indicates that persistent, long-term co-affiliations are associated with higher citation impact. Moreover, the findings highlight that an organisation's positional role within multi-affiliated publications can substantially influence its observed performance.

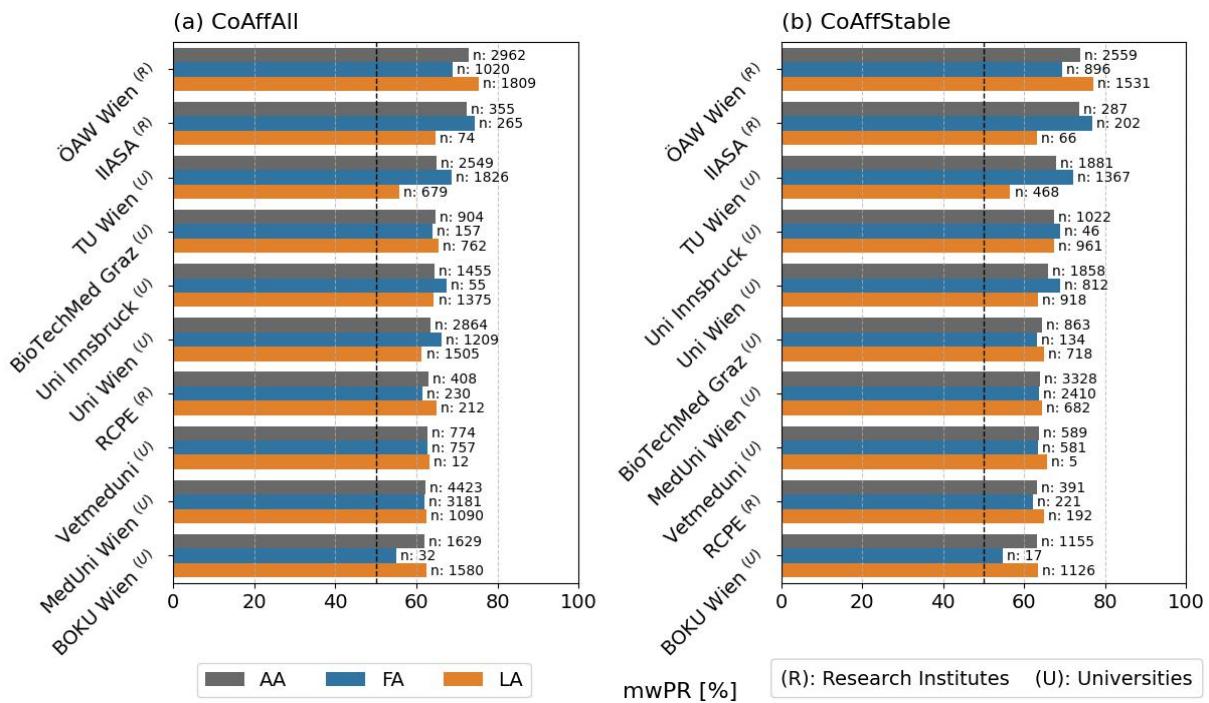

Figure 4: Citation impact of co-affiliated Austrian organisations. The top ten organisations, selected from CoAffAll with at least 300 records under the All-Affiliations (AA, no filter) scheme, are shown in both panels: (a) CoAffAll and (b) CoAffStable. Panels display mwPR(F) values for these organisations under the All-Affiliations (AA), First-Affiliation (FA), and Last-Affiliation (LA) filtering schemes. See for details on the link datasets and Table 4 for definitions of the affiliation-filtering schemes.
Organisational abbreviations: ÖAW Wien, Austrian Academy of Sciences (Vienna); IIASA, International Institute for Applied Systems Analysis; TU Wien, Vienna University of Technology; BioTechMed Graz, BioTechMed-Graz; Uni Innsbruck, University of Innsbruck; Uni Wien, University of Vienna; RCPE, Research Center Pharmaceutical Engineering; Vetmeduni, University of Veterinary Medicine Vienna; MedUni Wien, Medical University of Vienna; BOKU Wien, University of Natural Resources and Life Sciences Vienna.

Figure 5 illustrates the temporal evolution of mwPR for the top ten Austrian research organisations. ÖAW Wien consistently ranks as the top performer. In the CoAff–All dataset, it begins at 76% (n = 185) in 2013, fluctuates around the mid-70s between 2015 and 2019, and reaches 70% (n = 386) in 2022. The CoAff–Stable series yields even higher and smoother values (78% in 2013; 76% in 2018; 72% in 2022), indicating that persistent co-affiliations reinforce this citation advantage. BioTechMed Graz exhibits an extreme value in 2013 (98%), but this result is based on a single publication (n = 1). After 2016, its trajectory stabilises near the mid-60 % range in both datasets. IIASA exhibits peaks in mwPR at both the beginning and the end of the observation period. In the unfiltered CoAffAll dataset, mwPR is already high in 2013 (79%, *n* = 34) and rises again in 2021, reaching its maximum value (85%, *n* = 36). This late-period increase is preserved—in the CoAffStable dataset, where mwPR attains 85% in 2021, consistent with a concentration of high-impact output among persistent co-affiliations. Throughout the observation period, universities exhibit generally stable mwPR trajectories. Only BioTechMed Graz shows noticeable fluctuations, reflecting the sensitivity of citation-impact estimates in smaller publication portfolios.

Overall, the CoAffStable series (Figure 5b) yield higher mwPR values than the unfiltered CoAffAll series (Figure 5a). This pattern supports the interpretation that persistent, long-term co-affiliations are associated with greater and more stable citation impact, whereas short-lived or small-sample co-affiliations introduce volatility and can inflate single-year peaks.

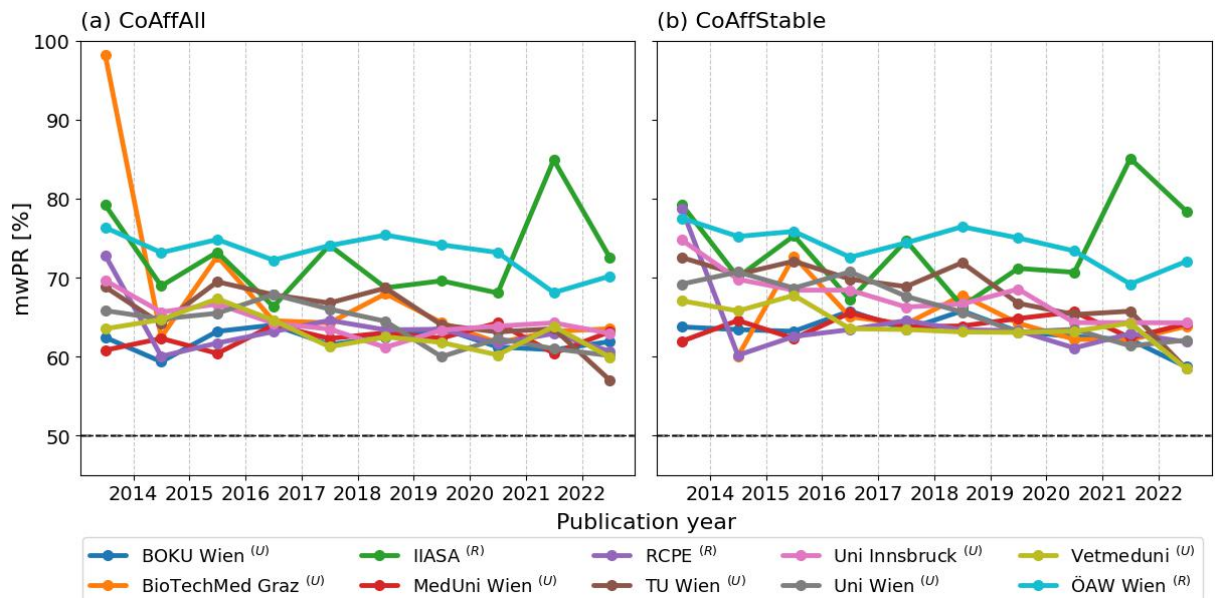

Figure 5: Same as Figure 3, but restricted to the ten organisations with at least 300 publication records under the All-Affiliations (AA) scheme. Organisational types are indicated as follows: (R) research institutes and (U) universities. Organisational abbreviations: ÖAW Wien, Austrian Academy of Sciences (Vienna); IIASA, International Institute for Applied Systems Analysis; TU Wien, Vienna University of Technology; BioTechMed Graz, BioTechMed-Graz; Uni Innsbruck, University of Innsbruck; Uni Wien, University of Vienna; RCPE, Research Center Pharmaceutical Engineering; Vetmeduni, University of Veterinary Medicine Vienna; MedUni Wien, Medical University of Vienna; BOKU Wien, University of Natural Resources and Life Sciences Vienna.

Figure 6 presents the organisational networks for CoAffAll and CoAffStable, analogous to Figure 1, and is restricted to the ten top-performing Austrian research organisations identified above based on mwPR values and a minimum threshold of 300 publications in the CoAff–All dataset. The networks were filtered to include only edges involving at least one of the ten top-performing Austrian organisations. Only nodes and edges with a co-affiliation strength of 50 or greater are shown. In contrast to Figure 1, which depicts the overall co-affiliation structure across all organisations above a higher edge-strength threshold of 250, Figure 6 focuses specifically on ego-networks centred on the top-performing organisations. This targeted filtering highlights their immediate organisational environments and reveals co-affiliation patterns that are less visible in the more comprehensive networks shown in Figure 1. In both datasets in Figure 6, the Austrian Academy of Sciences (ÖAW) Vienna exhibits its strongest connections with TU Wien and the University of Innsbruck. Universities are also interconnected with other universities and research organisations, indicating broader patterns of academic collaboration. The IIASE does not appear in either network, as it was excluded by the edge-strength threshold of 50, reflecting the comparatively small number of organisational connections and the limited sample size described above. The Medical University shows multiple links to universities and medical institutions (Figure 6, blue), highlighting its central role within the medical–academic collaboration landscape.

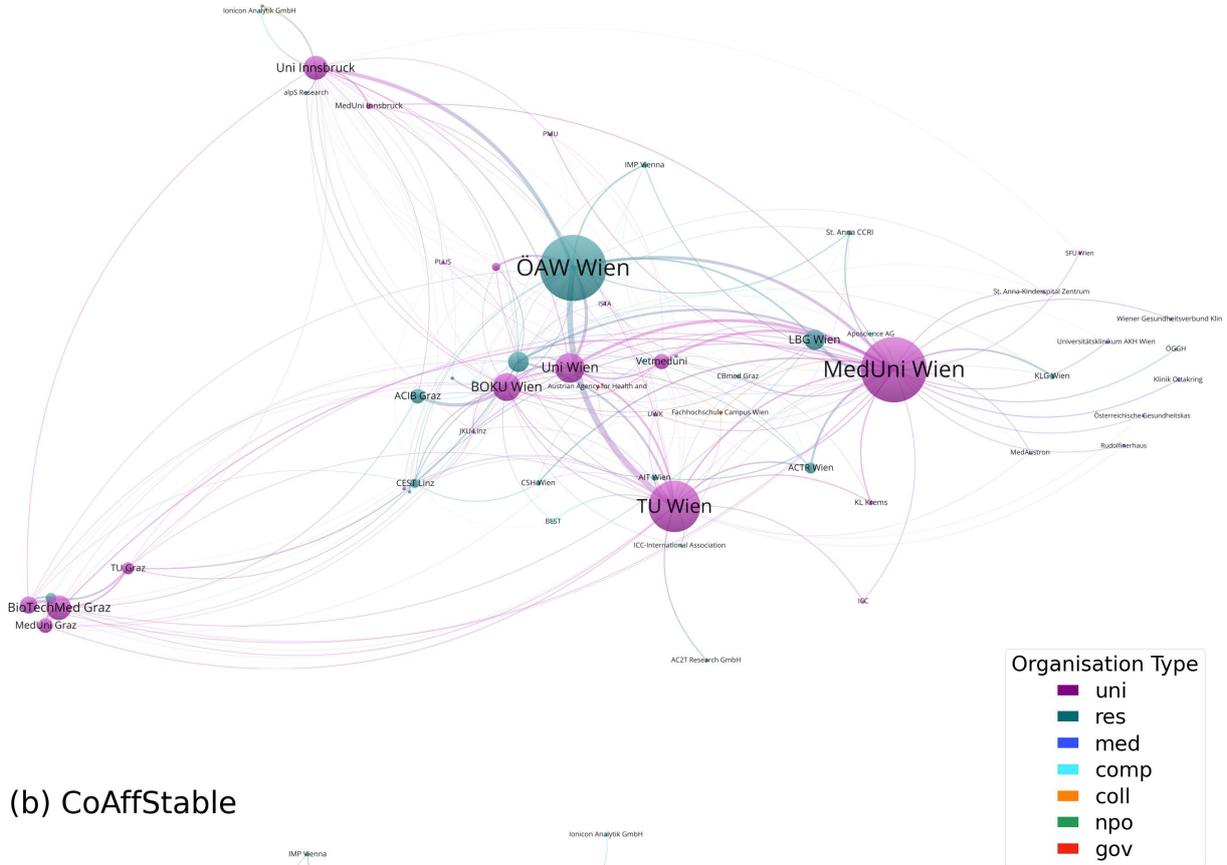

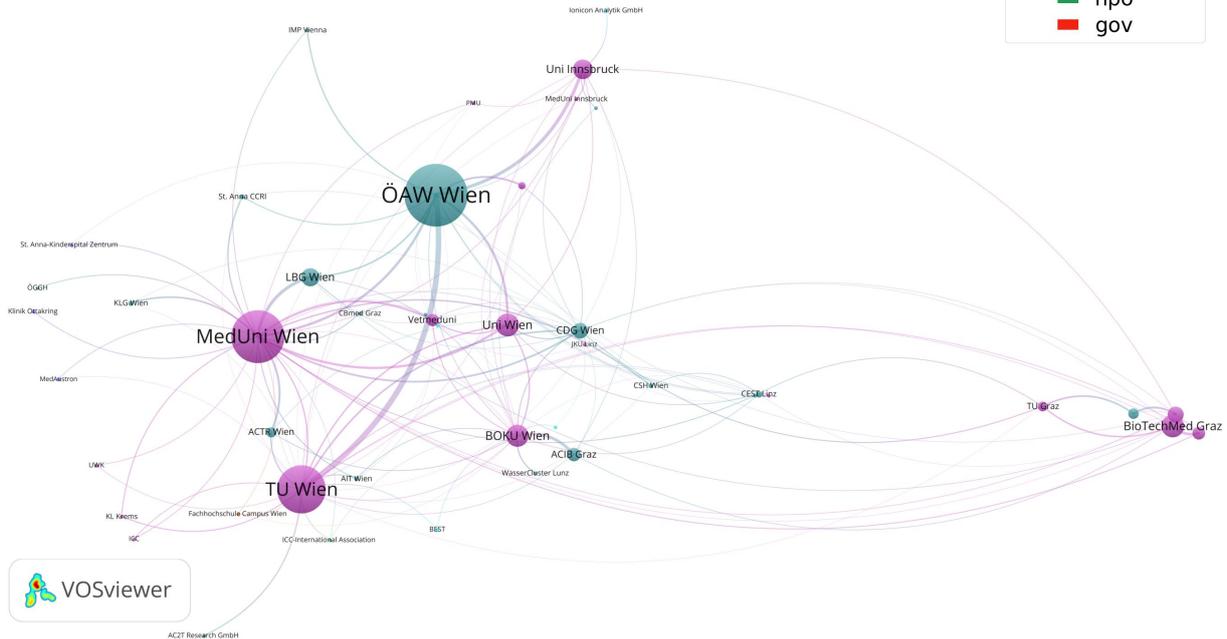

Figure 6: Organisational networks for CoAffAll and CoAffStable, equivalent to Figure 1, filtered to include only edges involving at least one of the top ten institutions and an edge-strength threshold of ≥50.

Organisations: universities (uni), research institutes (res), medical institutions (med), colleges (coll), companies (comp), government organisations (gov), and non-profit organisations (npo).

# 4   Conclusions

This study developed a scalable network-based framework for analysing multiple organisational co-affiliations that represents simultaneous affiliations held by individual researchers as relational links between organisations. The framework was applied to bibliometric data from Austria. The framework enabled a unified analysis of the structural organisation of co-affiliations and their implications for the allocation of organisational citation impact within a national research system. Using harmonised publication and affiliation metadata, two complementary co-affiliation networks were constructed for Austria. The first, CoAffAll, captures the complete set of observed simultaneous organisational affiliations, representing the full scope of co-affiliation activity. The second, CoAffStable, applies a conservative temporal filter, retaining only organisational pairs that recur in publications separated by at least two years. Analysing these networks in parallel allowed us to distinguish between transient and persistent co-affiliations. We assessed how short-lived and enduring organisational ties jointly shaped the structure of intra-organisational connections (between organisations of the same organisational type) and inter-organisational connections (between organisations of different types). This dual-dataset design extends previous studies that rely exclusively on stability-filtered affiliations and enables a more nuanced interpretation of organisational integration.

We estimated network regression models on both datasets to examine the determinants of co-affiliation ties. Model 1 focuses on intra-organisational co-affiliations (links between organisations of the same organisational type). Model 2 focuses on inter-organisational co-affiliations between different types. Across both models and datasets, geographical proximity emerged as a central structuring force: increasing travel time between organisations strongly and consistently reduced the likelihood and intensity of co-affiliation ties, underscoring the continued relevance of spatial constraints even in highly mobile research systems.

Beyond geography, the models revealed pronounced sectoral differences. Intra-organisational co-affiliations are strongly shaped by organisational scale and organisational role, with universities forming a dense and persistent core of shared appointments. While co-affiliations within medical institutions and colleges appear relatively frequently in the unfiltered network, these links largely dissipate under temporal filtering, suggesting that many reflect short-term or role-specific arrangements rather than durable organisational ties. Co-affiliations within the private-sector, government and non-profit organisations remained rare and weakly institutionalised across both datasets. In the inter-organisational model, temporal filtering substantially reduced cross-sector ties, isolating university–research institute co-affiliations as uniquely resilient. This pattern points to structurally embedded organisational integration rather than episodic collaboration. The results highlight the selective nature of durable inter-organisational relationships.

In addition to mapping co-affiliation structures, we assessed how multiple concurrent affiliations shaped the allocation of citation impact across organisations and organisational types. Citation impact was measured using field- and year-normalised citation scores based on weighted Hazen percentiles, aggregated as mwPR. Fractional counting ensured that publications by multi-affiliated authors contribute proportionally to all affiliated organisations and organisational types. Sensitivity analyses using alternative affiliation-filtering schemes and minimum publication thresholds demonstrated that citation-impact estimates and organisational rankings are sensitive to how affiliations are represented and to sample-size constraints. Nonetheless, robust patterns persist across all specifications.

Across organisational types, research institutes and universities consistently exhibited the highest citation impact, with research institutes outperforming other sectors. Although absolute mwPR values vary depending on whether first-, all-, or last-listed affiliations were retained, the relative ordering of organisational types remained stable. Persistent co-affiliations captured in the CoAffStable dataset are systematically associated with higher and more stable citation impact, indicating that enduring institutional ties coincide with greater scientific visibility. At the organisational level, publication-volume thresholds substantially stabilised rankings by reducing small-sample inflation, while leading organisations retained their positions across datasets and filtering schemes.

Taken together, these findings demonstrate that multiple organisational affiliations are a structured and consequential feature of contemporary research systems. By integrating network-based representations of co-affiliations with temporally sensitive filtering and fractional citation-impact measures, this study provides a robust framework for analysing organisational collaboration and organisational performance. More broadly, our results show that the modelling of multi-affiliated researchers has material consequences for both the observed structure of research systems and the attribution of scientific impact.

## Code availability

All code used for data processing, network construction, and model fitting in this study is openly available. The complete analysis pipeline, including source code, containerised workflows, and configuration files, is hosted on GitHub https://gitlab.tuwien.ac.at/metalab/co-affiliation-analyses).

The software used in this study has been archived on Zenodo and assigned a persistent DOI (Co-Affiliation Analysis, CAA, version v0.9.2; https://doi.org/10.5281/zenodo.18088962). All results reported in this paper were generated using this archived version of the code [26]. To support computational reproducibility, a pre-built Docker image containing the full software environment and all command-line entry points used in the analyses are available via Docker Hub (https://hub.docker.com/repository/docker/metalabvienna/co-affiliation-network). The Docker image version v0.9.2 corresponds to the Zenodo-archived code and the Git repository tag v0.9.2. Detailed instructions for reproducing the analyses—either by running the scripts directly or by using the Docker container—are provided in the repository documentation.

## Data availability

Bibliometric data were derived from Scopus® data (Elsevier) accessed via the Scopus APIs. Scopus® is a registered trademark of Elsevier B.V. Further bibliometric data were derived from a bibliometric in-house database of the Max Planck Society (MPG). This database was developed and is maintained in cooperation with the Information Retrieval Service of the CPT Section (IVS-CPT) at the Max Planck Institute for Solid State Research. The database is derived from the Scopus database via the German "Kompetenznetzwerk Bibliometrie" (see https://bibliometrie.info/en/about-kb/) funded by the German Federal Ministry of Research, Technology and Space (grant 16WIK2101A).

The input data used in the processing pipeline are publicly available on Zenodo. The Scopus-derived publication-level dataset, comprising bibliographic records of journal articles published between 2013 and 2023 in which at least one author reports two or more simultaneous institutional affiliations located in Austria, is available under the DOI https://doi.org/10.5281/zenodo.17953805 [27]. Affiliation-level metadata, including harmonised parent-organisation assignments encoded in the variable *affiliation_id_parent* and geographic coordinates derived from affiliation addresses, are available under the DOI https://doi.org/10.5281/zenodo.17953805 [28]. Travel-time estimates between institutional affiliations were computed as part of the analysis using geocoded affiliation locations and routing based on OpenStreetMap data via the open-source Valhalla routing engine are provided under the DOI https://doi.org/10.5281/zenodo.17954105 [29].

## Author Contributions

Software development: Christoph Schlager.
Methodology: Christoph Schlager, Lutz Bornmann, Gerald Schweiger.
Conceptualization: Christoph Schlager, Lutz Bornmann, Gerald Schweiger.
Data curation: Christoph Schlager, Lutz Bornmann.
Formal analysis: Christoph Schlager.
Visualization: Christoph Schlager.
Writing – original draft: Christoph Schlager, Lutz Bornmann, Gerald Schweiger.
Writing – review & editing: Christoph Schlager, Lutz Bornmann, Gerald Schweiger.

# References

[1]  Yegros-Yegros A, Capponi G, Frenken K. A spatial-institutional analysis of researchers with multiple affiliations. PLOS ONE 2021;16:e0253462. https://doi.org/10.1371/journal.pone.0253462.
[2]  Hottenrott H, Lawson C. A first look at multiple institutional affiliations: a study of authors in Germany, Japan and the UK. Scientometrics 2017;111:285–95. https://doi.org/10.1007/s11192-017-2257-6.
[3]  Meho LI, Akl EA. Using bibliometrics to detect questionable authorship and affiliation practices and their impact on global research metrics: a case study of 14 universities. Quant Sci Stud 2025;6:63–98. https://doi.org/10.1162/qss_a_00339.
[4]  Huang M-H, Chang Y-W. Multi-institutional authorship in genetics and high-energy physics. Phys Stat Mech Its Appl 2018;505:549–58. https://doi.org/10.1016/j.physa.2018.03.091.
[5]  Katz JS, Martin BR. What is research collaboration? Res Policy 1997;26:1–18. https://doi.org/10.1016/S0048-7333(96)00917-1.
[6]  Hottenrott H, Lawson C. What is behind multiple institutional affiliations in academia? Sci Public Policy 2022;49:382–402. https://doi.org/10.1093/scipol/scab086.
[7]  Hottenrott H, Rose ME, Lawson C. The rise of multiple institutional affiliations in academia. J Assoc Inf Sci Technol 2021;72:1039–58. https://doi.org/10.1002/asi.24472.
[8]  Way SF, Morgan AC, Larremore DB, Clauset A. Productivity, prominence, and the effects of academic environment. Proc Natl Acad Sci 2019;116:10729–33. https://doi.org/10.1073/pnas.1817431116.
[9]  Lin C, Huang M, Chen D. The inter-institutional and intra-institutional multi-affiliation authorships in the scientific papers produced by the well-ranked universities. J Informetr 2025;19:101635. https://doi.org/10.1016/j.joi.2024.101635.
[10] Tong S, Yue T, Shen Z, Yang L. The effect of national and international multiple affiliations on citation impact 2020. https://doi.org/10.48550/arXiv.2001.06803.
[11] Halevi G, Rogers G, Guerrero-Bote VP, De-Moya-Anegón FD-M-A. Multi-affiliation: a growing problem of scientific integrity. El Prof Inf 2023:e320401. https://doi.org/10.3145/epi.2023.jul.01.
[12] Sanfilippo P, Hewitt AW, Mackey DA. Plurality in multi-disciplinary research: multiple institutional affiliations are associated with increased citations. PeerJ 2018;6:e5664. https://doi.org/10.7717/peerj.5664.
[13] Elsevier. Scopus — Abstract and Citation Database 2025. https://www.scopus.com (accessed November 17, 2025).
[14] Burger M, van Oort F, Linders G-J. On the specification of the gravity model of trade: zeros, excess zeros and zero-inflated estimation. Spat Econ Anal 2009;4:167–90. https://doi.org/10.1080/17421770902834327.
[15] Breschi S, Lissoni F. Mobility of skilled workers and co-invention networks: an anatomy of localized knowledge flows. J Econ Geogr 2009;9:439–68. https://doi.org/10.1093/jeg/lbp008.
[16] Hoekman J, Frenken K, van Oort F. The geography of collaborative knowledge production in Europe. Ann Reg Sci 2009;43:721–38. https://doi.org/10.1007/s00168-008-0252-9.
[17] Ponds R, van Oort F, Frenken K. The geographical and institutional proximity of research collaboration. Pap Reg Sci 2007;86:423–44. https://doi.org/10.1111/j.1435-5957.2007.00126.x.
[18] Vedres B, Stark D. Structural folds: generative disruption in overlapping groups. Am J Sociol 2010;115:1150–90. https://doi.org/10.1086/649497.
[19] Valhalla contributors. valhalla/valhalla 2025.
[20] Zheng YY, Shida Y, Takayasu H, Takayasu M. Enhancing the gravity model for commuters: time-and-spatial-structure-based improvements in Japan's metropolitan areas. PLOS ONE 2025;20:e0329603. https://doi.org/10.1371/journal.pone.0329603.
[21] Bornmann L, Williams R. An evaluation of percentile measures of citation impact, and a proposal for making them better. Scientometrics 2020;124:1457–78. https://doi.org/10.1007/s11192-020-03512-7.
[22] Waltman L, van Eck NJ. Field-normalized citation impact indicators and the choice of an appropriate counting method. J Informetr 2015;9:872–94. https://doi.org/10.1016/j.joi.2015.08.001.
[23] Waltman L, van Eck NJ, van Leeuwen TN, Visser MS, van Raan AFJ. Towards a new crown indicator: an empirical analysis. Scientometrics 2011;87:467–81. https://doi.org/10.1007/s11192-011-0354-5.
[24] Studies (CWTS) C for S and T. CWTS Leiden Ranking Traditional Edition. CWTS Leiden Rank Tradit Ed n.d. https://traditional.leidenranking.com (accessed December 2, 2025).
[25] Elizondo AR, Calero-Medina C, Visser MS. The three-step workflow: a pragmatic approach to allocating academic hospitals' affiliations for bibliometric purposes. J Data Inf Sci 2022;7:20–36. https://doi.org/10.2478/jdis-2022-0006.
[26] Schlager C. Co-affiliation analysis (CAA) 2025. https://doi.org/10.5281/zenodo.17972957.
[27] Schlager C. Scopus dataset of co-affiliated research publications in Austria (2013–2023) 2025. https://doi.org/10.5281/zenodo.17952177.
[28] Schlager C. Scopus dataset of co-affiliation metadata for research publications in Austria (2013–2023) 2025. https://doi.org/10.5281/zenodo.18082675.
[29] Schlager C. Travel-time enriched affiliation pair dataset for the scopus co-affiliation metadata (Austria, 2013–2023) 2025. https://doi.org/10.5281/zenodo.18082558.

# Appendix

The Appendix shows the results of the organisational mwPR rankings sensitivity to alternative minimum publication thresholds. Raising the minimum publication threshold from 300 to 400 led to a modest but informative adjustment in the composition of the top-ranked organisations (Figure A1). The International Institute for Applied Systems Analysis (IIASA) no longer met the inclusion criterion, reflecting its comparatively small publication portfolio, while the Ludwig Boltzmann Society (LBG) Vienna entered the top ten institutions across all attribution schemes. Despite these changes, the overall ranking structure remained stable: ÖAW Wien consistently led across schemes and datasets, followed by a largely unchanged group of major universities and research-performing organisations. Together, these findings indicate that the principal results are robust to moderate increases in the minimum sample-size requirement, while illustrating how stricter thresholds began to exclude more specialised organisations with limited publication output.

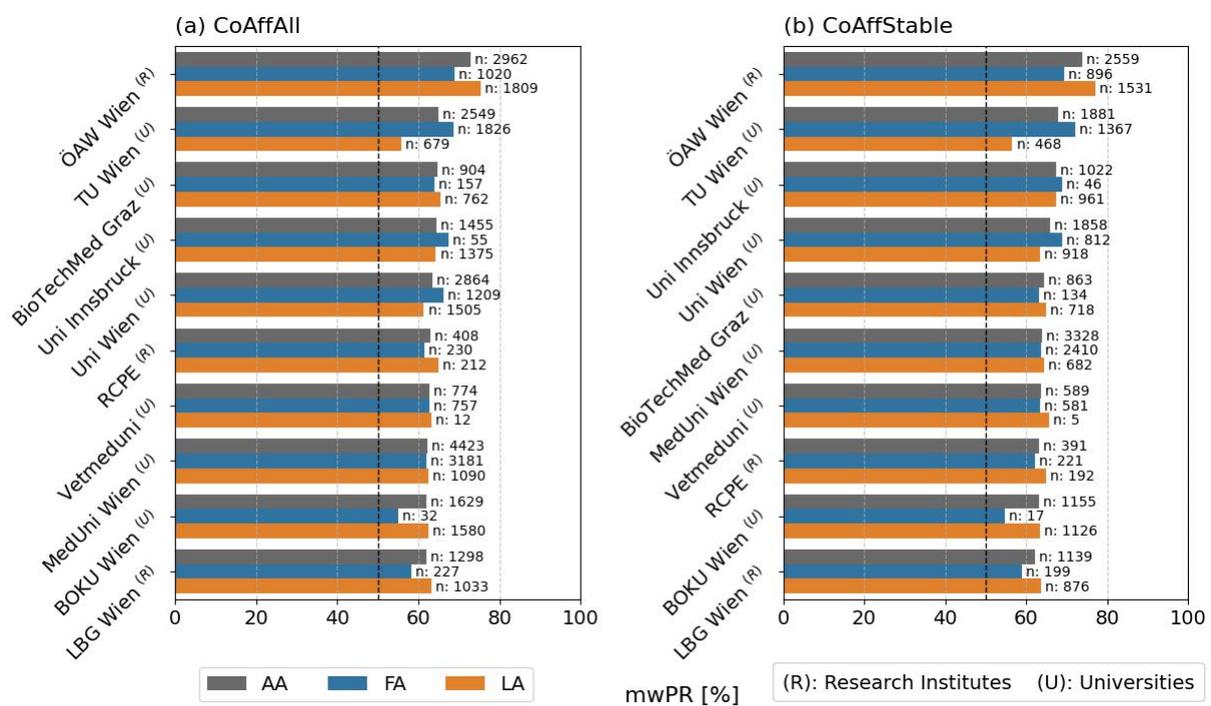

Figure A1: Citation impact of co-affiliated Austrian organisations. The top ten organisations, selected from CoAffAll with at least 400 records under the All-Affiliations (AA, no filter) scheme, are shown in both panels: (a) CoAffAll and (b) CoAffStable. Panels display mwPR(F) values for these organisations under the All-Affiliations (AA), First-Affiliation (FA), and Last-Affiliation (LA) filtering schemes. See Table 2 for details on the link datasets and Table 4 for definitions of the affiliation-filtering schemes.
Organisational abbreviations: ÖAW Wien, Austrian Academy of Sciences (Vienna); IIASA, International Institute for Applied Systems Analysis; TU Wien, Vienna University of Technology; BioTechMed Graz, BioTechMed-Graz; Uni Innsbruck, University of Innsbruck; Uni Wien, University of Vienna; RCPE, Research Center Pharmaceutical Engineering; Vetmeduni, University of Veterinary Medicine Vienna; MedUni Wien, Medical University of Vienna; BOKU Wien, University of Natural Resources and Life Sciences Vienna.

Increasing the minimum publication threshold to 600 yielded a more selective ranking that further concentrated on organisations with large and sustained publication output (Figure A2). Relative to the 400-threshold setting, the composition of the top ten narrowed, as smaller or more specialised research organisations are progressively excluded, while the relative ordering of the remaining organisations remained largely unchanged. ÖAW Wien continued to lead across all attribution schemes and both datasets, followed by a stable group of major universities. Research-focused organisations with intermediate publication volumes, such as the Christian Doppler Gesellschaft (CDG) Wien, remained represented, whereas organisations near the minimum threshold displayed larger shifts in mwPR across attribution schemes and dataset filtering. Despite these compositional changes, mwPR values varied only modestly across thresholds, and the systematic increase observed when moving from unfiltered to stable co-affiliation structures persisted. Together, these findings indicate that the principal results remained robust under substantially stricter sample-size requirements, while illustrating how higher thresholds increasingly privilege organisations with extensive and persistent publication activity.

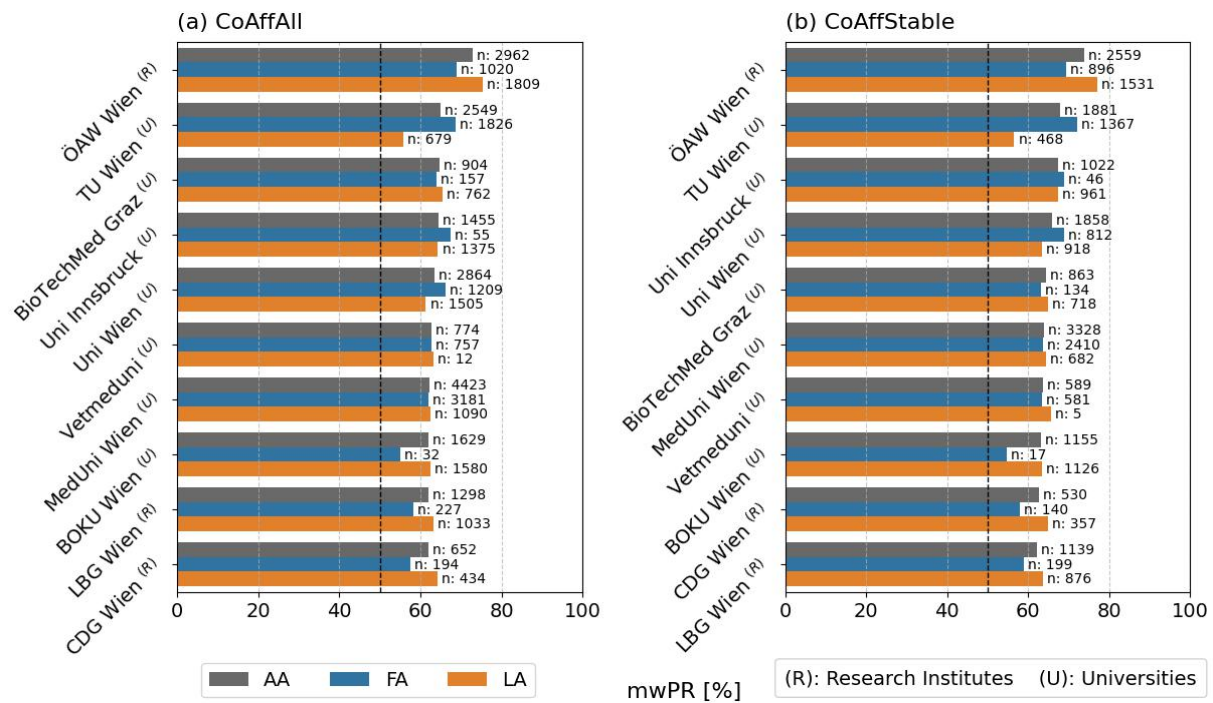

Figure A2: Citation impact of co-affiliated Austrian organisations. The top ten organisations, selected from CoAffAll with at least 600 records under the All-Affiliations (AA, no filter) scheme, are shown in both panels: (a) CoAffAll and (b) CoAffStable. Panels display mwPR(F) values for these organisations under the All-Affiliations (AA), First-Affiliation (FA), and Last-Affiliation (LA) filtering schemes. See Table 2 for details on the link datasets and Table 4 for definitions of the affiliation-filtering schemes.

Organisational abbreviations: ÖAW Wien, Austrian Academy of Sciences (Vienna); TU Wien, Vienna University of Technology; BioTechMed Graz, BioTechMed-Graz; Uni Innsbruck, University of Innsbruck; Uni Wien, University of Vienna; RCPE, Research Center Pharmaceutical Engineering; Vetmeduni, University of Veterinary Medicine Vienna; MedUni Wien, Medical University of Vienna; BOKU Wien, University of Natural Resources and Life Sciences Vienna; CDG Wien, Christian Doppler Research Association.